\documentstyle[12pt]{article}
\begin{document}
\begin{center}
{\LARGE Resonance model for $\pi$N scattering and \\ $\eta$-meson production
in the S$_{11}$ channel}\\[5mm]
Ch.~Sauermann \footnote[1]
{e-mail:sauerman@tpri6f.gsi.de}, B.L.~Friman and W.~N\"orenberg\\[3mm]
GSI, Postfach 11 05 52, D-64220 Darmstadt, Germany\\
and\\
Institut f\"ur Kernphysik, Technische Hochschule Darmstadt,\\
D-64289 Darmstadt, Germany\\[5mm]

\end{center}

\section*{Abstract}
A model for $\pi$N scattering and $\eta$-meson production in the
S$_{11}$ channel
is presented. The model includes $\pi$N-scattering Born terms as well as the
N$^*${} resonances
S$_{11}$(1535) and S$_{11}$(1610). The $T$-matrix is computed in
the $K$-matrix approximation. The parameters of the model are determined by
fitting the elastic $\pi$N-scattering $T$-matrix to empirical data. We find
an excellent fit for all energies up to $\sqrt{s}$ = 1.75 GeV.
Furthermore,
a good description of the cross section for $\pi^- + p \rightarrow \eta + n$
is obtained without further adjustment of parameters.

\newpage

\section{Introduction}

During the last few years there has been renewed interest in the physics of
the $\eta$-meson. New data on $\eta$-production in different reactions are
becoming available. For example, the photoproduction of $\eta$-mesons is
being studied at MAMI \cite{1} and ELSA \cite{1a}, while $\eta$-production in
hadronic and heavy-ion collisions is under investigation at SATURNE
\cite{1b} and GSI \cite{1c}, respectively. Calculations of these processes
have been performed by several groups (see e.g.~refs.~[5-9]).

The ultimate aim of our investigation is a better understanding of the
$\eta$-properties in an hadronic environment and the possible consequences
for $\eta$-production in heavy-ion collisions. However, before attacking the
relatively complex problem of medium effects, we need a better understanding
of more elementary processes, like e.g. $\eta$-production in hadronic
collisions and in photo-induced reactions. New accurate data on these
processes will in the near future provide very useful constraints on models
for the elementary $\eta$-meson--hadron interactions and the properties of the
$\eta$-meson at normal
nuclear matter density and below. The investigation of possible $\eta$-meson
bound
states in nuclei could also yield information on $\eta$-meson--nucleon
interactions
at low densities \cite {7}. Since the $\eta$-meson couples strongly to the
S$_{11}$(1535) resonance, the in-medium properties of the $\eta$-meson and
this resonance are strongly interdependent. Thus, $\eta$-production on nuclei
also probes the in-medium properties of the S$_{11}$(1535) resonance.

A useful framework for the study of hadrons in nuclear matter is an
effective field theory with hadronic degrees of freedom. In such theories
one needs as input the hadronic coupling constants.
It is the aim of this
letter to construct a model for pion-induced $\eta$-production on the nucleon.
The parameters, in particular the $\eta$NN$^*${} coupling
constant, are determined by fitting data on elastic $\pi$N scattering.
The consistency of our model is then
checked by comparing the cross section for the inelastic channel
$\pi$N$\rightarrow \eta$N with experiment. The present calculation
is a first step in a systematic investigation of $\eta$-interactions with
hadrons and hadronic matter, as outlined above.

\section{Model lagrangian}

Pion-nucleon scattering in the S$_{11}$ channel at center-of-mass (c.m.)
energies below $\sqrt{s}=$1.8~GeV involves two N$^*$ resonances \cite{8,9}.
The lower one, the S$_{11}$(1535) resonance, decays into $\pi$N, $\eta$N and
$\pi\pi$N with the branching ratios 35--55$\,\%$, 30--50$\,\%${} and
5--20$\,\%$, respectively \cite{10}. The upper one, i.e. the
S$_{11}$(1650), couples strongly to the $\pi$N and $\pi\pi$N channels with
branching ratios of 60--80$\,\%${} and 5--20$\,\%$, respectively, but only
weakly to the $\eta$N channel with a probability of approximately 1$\,\%$.

In order to describe these experimental facts we include the
following interaction terms in the lagrangian:

\begin{eqnarray}\label{eq:a}
{\cal L}_I &=&-i  g_{\pi \mbox{\scriptsize NN}}\overline{\Psi}_
{\mbox{\scriptsize N}} \gamma_5\vec{\tau}\Psi_{\mbox{\scriptsize
N}}\vec{\pi}
-g_{\sigma \mbox{\scriptsize NN}} \overline{\Psi}_
{\mbox{\scriptsize N}} \Psi_{\mbox{\scriptsize
N}}\sigma
-g_{\sigma\pi\pi} \vec{\pi}^2\sigma  \nonumber \\
& &-i g_{\pi \mbox{\scriptsize N}\mbox{\scriptsize N}^*_1}\overline{\Psi}_
{\mbox{\scriptsize N}^*_1}\vec{\tau}\Psi_{\mbox{\scriptsize
N}}\vec{\pi}+\mbox{h.c.}
-i g_{\pi \mbox{\scriptsize N}\mbox{\scriptsize N}^*_2}\overline{\Psi}_
{\mbox{\scriptsize N}^*_2}\vec{\tau}\Psi_{\mbox{\scriptsize
N}}\vec{\pi}+\mbox{h.c.} \\
& &-i g_{\eta \mbox{\scriptsize N}\mbox{\scriptsize N}^*_1}\overline{\Psi}_
{\mbox{\scriptsize N}^*_1}\Psi_{\mbox{\scriptsize
N}}\eta+\mbox{h.c.}  \nonumber \\
& &-i g_{\zeta \mbox{\scriptsize N}\mbox{\scriptsize N}^*_1}\overline{\Psi}_
{\mbox{\scriptsize N}^*_1}\gamma_5\Psi_{\mbox{\scriptsize
N}}\zeta+\mbox{h.c.}
-i g_{\zeta \mbox{\scriptsize N}\mbox{\scriptsize N}^*_2}\overline{\Psi}_
{\mbox{\scriptsize N}^*_2}\gamma_5\Psi_{\mbox{\scriptsize
N}}\zeta +\mbox{h.c.} \nonumber.
\end{eqnarray}
The first line contains the interaction terms of the
linear sigma-model \cite{12,13} which is chirally invariant and describes
low-energy $\pi$N scattering reasonably well. In this model
the following relations between the coupling constants hold:
$g_{\sigma\mbox{\scriptsize NN}}=g_{\pi\mbox{\scriptsize NN}}${} and
$g_{\sigma\pi\pi}=g_{\pi\mbox{\scriptsize NN}}(m_{\sigma}^2-m_{\pi}^2)/
2m_{\mbox{\scriptsize N}}$, where $m_{\sigma}, m_{\pi}${} and
$m_{\mbox{\scriptsize N}}${} denote the masses of  $\sigma$, $\pi$ and
nucleon, respectively. The interaction terms in the second line
describe the $\pi$NN$^*${} interaction for the two
N$^*$ resonances (N$^*_1\equiv$S$_{11}$(1535), N$^*_2\equiv$S$_{11}$(1650))
whereas those in the third line account for the $\eta$NN$^*_1${} coupling.
The weak $\eta$NN$^*_2${} and $\eta$NN{} couplings are neglected \cite{11}.

In principle, one could attempt to describe the $\pi\pi$N{} decay of the
S$_{11}${} resonances by the two-step process N$^* \rightarrow
\sigma\mbox{N}\rightarrow \pi\pi$N. This would require only two additional
coupling terms of the resonances to the $\sigma$N channel. The
subsequent decay of the $\sigma${} into two pions would be described by the
third term in (1). However, there are other intermediate states, like
$\rho$N and $\pi\Delta$, which also contribute to the $\pi\pi$N decay
channel and should be handled on the same footing as the $\sigma$N
process. Thus, a proper treatment of this decay channel, which plays a somewhat
peripheral role in our investigation, would be unduly complex. Consequently,
we choose a phenomenological approach, where the physical two-pion
continuum is parametrized by means of an effective scalar field $\zeta${} of
positive parity, mass m$_{\zeta}$=400 MeV and zero width. The field $\zeta$
interacts only through the $\zeta$NN$^*$ couplings, which are
given in the last line of (\ref{eq:a}).

\section{$T$-matrix}

The $T$-matrix is related to the $K$-matrix by the integral equation
\begin{equation}\label{eq:d}
T=K-i\pi K\delta(E-H_0)T,
\end{equation}
where $H_0${} describes the free motion of the two particles in one of the
three coupled channels $\pi$N{}, $\eta$N{} and $\zeta$N. The
$K$-matrix can be derived from
the lagrangian by solving another, more
complicated, integral equation. However, since any approximation, which
leaves the $K$-matrix hermitian, yields a unitary $S$-matrix \cite{11a}, we
circumvent this step by identifying the $K$-matrix elements with the
diagrams shown in fig.~1 (c.f.~ref.~\cite{4a}). In the tree approximation
to the linear $\sigma$-model the direct and crossed nucleon pole terms and
the $\sigma$-exchange term are included \cite{12,13}. These terms, the so
called Born terms, correspond to the first three diagrams of $K_{\pi\pi}$ in
fig.~1. We also include the direct contributions from the resonances. The
corresponding crossed diagrams are suppressed due to the large masses in
intermediate states.

In the diagrams for the $K$-matrix the intermediate state
particles propagate with their physical masses but with zero widths; e.g.,
for the S$_{11}$(1535) resonance in the intermediate state the corresponding
inverse propagator is given by
$S(p)^{-1}= p\!\!\!/-m_1$, where
$m_1$ is real. The summation of the processes implied by eq.~(\ref{eq:d})
generates the imaginary parts needed to satisfy unitarity, and consequently
also the correct resonance widths.
In this approximation the $K$-matrix, defined by the diagrams in fig.~1, is
obviously free of divergences. Since the intermediate meson-nucleon states in
eq.~(2) are restricted to be on-shell, this also applies to the $T$-matrix.
Consequently, there is no need to renormalize e.g.~the coupling constants.
Furthermore, since the integral equation
(\ref{eq:d}) reduces to an algebraic one after projection on partial waves,
one can, given a suitable approximation to the $K$-matrix, write
down the $T$-matrix analytically.

We introduce form factors for the off-shell nucleon at the $\pi$NN vertex
\cite{14}
\begin{equation}\label{eq:f}
F=\frac{\Lambda^4}{\Lambda^4+(p_{\mbox{\scriptsize N}}^2
-m_{\mbox{\scriptsize N}}^2)^2}
\end{equation}
and for the off-shell $\sigma$-meson at the $\sigma\pi\pi${} and
$\sigma$NN vertices
\begin{equation}\label{eq:f2}
G_{1,2}=\frac{\lambda_{1,2}^2}
{\lambda_{1,2}^2-p_{\sigma}^2} .
\end{equation}
Here $p_{\mbox{\scriptsize N}}$ and $p_{\sigma}$ are the 4-momenta of the
off-shell nucleon and $\sigma$-meson, respectively.

The $T$-matrix in the S$_{11}$ channel is obtained by
projecting eq.~(\ref{eq:d})  onto total angular momentum $j=1/2$,
isospin $t=1/2$ and
negative parity. Since the $\pi$- and $\eta$-mesons have negative parity,
only $l=0${} for the relative angular momentum is allowed in the
$\pi$N and $\eta$N intermediate states.
On the other hand, the $\zeta$N intermediate state
must be p-wave, since the $\zeta$-meson has positive parity.
With our choice of form factors the projection can be done analytically.

The resulting $T$-matrix can be given in closed form. However, since the full
expression, keeping all three channels is quite lengthy,
we do not present it here. Instead, we show the solution for
the simpler case, where the $\pi\pi$N channel is neglected. For elastic
$\pi$N scattering in the S$_{11}$ channel we then find
\begin{equation}\label{eq:g}
T_{\pi\pi}^0= \frac{-i m_{\mbox{\scriptsize N}} q} {(4\pi)^2\sqrt{s}}
\left[\frac{-\left[F(k)\gamma-v\right]\left[K^B w
+\beta\right]+\alpha w}{\left[F(k)\gamma-v\right)]
\left[F(q)(K^B w+\beta)-w\right]-\alpha w F(q)}\right] ,
\end{equation}
where $\sqrt{s}${} is the invariant mass in the $\pi$N channel.
The absolute values of the relative three-momenta in the
c.m.-system of the  $\pi$N and $\eta$N system are denoted by $q${}
and $k$, respectively.
Furthermore, $v=\sqrt{s}-m_1$,
$w=\sqrt{s}-m_2$, where $m_1${} and $m_2$ denote the masses of the two
resonances and $F(k)=m_{\mbox{\scriptsize N}}
k/\left(4(2\pi^2)\sqrt{s}\right)$.
The other quantities are given by
\begin{eqnarray}\label{eq:h}
\alpha&=&\frac{-6i\pi g_{\pi\mbox{\scriptsize N}\mbox{\scriptsize N}^*_1}^2}
{m_{\mbox{\scriptsize N}}}
(E_q+m_{\mbox{\scriptsize N}}), \
\beta=\frac{-6i\pi g_{\pi\mbox{\scriptsize N}\mbox{\scriptsize N}^*_2}^2}
{m_{\mbox{\scriptsize N}}}
(E_q+m_{\mbox{\scriptsize N}}) ,\\
\gamma&=&\frac{-2i\pi g_{\eta\mbox{\scriptsize N}\mbox{\scriptsize N}^*_1}^2}
{m_{\mbox{\scriptsize N}}}
(E_k+m_{\mbox{\scriptsize N}}) \nonumber
\end{eqnarray}
where $E_q=\sqrt{q^2+m_{\mbox{\scriptsize N}}^2}\,\,${} and
$E_k=\sqrt{k^2+m_{\mbox{\scriptsize N}}^2}$.
The projection of the non-resonant
$\pi$N Born terms onto the S$_{11}$ channel is denoted by $K^B$.
In order to obtain the empirical scattering lengths in the tree
approximation to the linear $\sigma$-model, the $\pi$NN coupling constant
must be renormalized \cite{13}. This renormalization, which effectively
accounts for loop corrections, is usually done by fixing the axial
coupling constant to its empirical value $g_A \approx 1.3$.
We choose a pragmatic approach and adjust the effective $\pi$NN coupling
constant to the
s-wave $\pi$N scattering length $a_1${} for isospin=1 in the t-channel,
where the $\sigma$-exchange term does not contribute.
The mass of the $\sigma$-meson
is then fixed by
the isospin=0 scattering length $a_0$.
The remaining five coupling constants as well as the masses
of the two resonances are then determined by
fitting the elastic $\pi$N-scattering data up to $\sqrt{s} = $ 1.84 GeV.

\section{Elastic $\pi$N-scattering data}
There are several partial-wave analyses of the $\pi$N-scattering data
available. We employ the solutions obtained by H\"{o}hler and the
Karlsruhe-Helsinki group \cite{8} and by Cutkosky and the CMU-LBL group
\cite{9}.

In our least-square fit we find a fairly shallow, but nevertheless clear
minimum.
The resulting parameter sets are given in table 1 in rows 1 and 2.

\begin{center}
\parbox[t]{13cm}{
Table 1. Parameter values obtained by fitting the Karlsruhe-Helsinki (KH)
and CMU-LBL solutions.}\\[0.3cm]
\begin{tabular}{|l||l|l|l|l|l|l|l|}\hline
& $m_1$ [GeV] & $m_2$ [GeV]& $g_{\pi\mbox{\scriptsize N}\mbox{\scriptsize
N}^*_1}$ &
$g_{\pi\mbox{\scriptsize N}\mbox{\scriptsize N}^*_2}$ &
$g_{\eta\mbox{\scriptsize N}\mbox{\scriptsize N}^*_1}$ &
$g_{\zeta\mbox{\scriptsize N}\mbox{\scriptsize N}^*_1}$ &
$g_{\zeta\mbox{\scriptsize N}\mbox{\scriptsize N}^*_2}$ \\ \hline\hline
KH & 1.55 & 1.70 & 0.67 & 1.17 & 2.10 & 2.36 & 4.85 \\ \hline
CMU-LBL &1.53  & 1.70  & 0.64  & 1.26  & 2.13 & 1.89 &  4.37 \\ \hline
KH2 & 1.55 & 1.69 & 0.70 & 1.20 & 2.16 & 3.08 & 5.00 \\ \hline
\end{tabular}
\end{center}
\vspace{0.5cm}
The $\pi$NN coupling constant
$g_{\pi\mbox{\scriptsize N}\mbox{\scriptsize N}}$=8.99 which corresponds to
$g_A = 1.48$,
and the $\sigma$-meson mass $m_{\sigma}$=640 MeV
are fixed by
fitting the scattering lengths, as described above.
The cutoff masses
$\Lambda=3.1${} GeV, $\lambda_1=2.0${} GeV and $\lambda_2=2.1${} GeV
 are determined by fitting the data below the first resonance.

We note that the two fits are of similar quality and yield very similar
parameter sets.  In figs.~2 and 3 we show the fit to the Karlsruhe-Helsinki
solution which is well reproduced up to and including the energy range
of the second resonance. The discrepancy at energies beyond $\sqrt{s} $=
1.75 GeV is most likely due to other resonances, not included in our model.

Since the presence of form factors complicates the question of gauge
invariance e.g.~in the process  $\gamma +p \rightarrow \eta+p$,
we have also made a fit to the Karlsruhe-Helsinki solution without
a form factor at the $\pi$NN vertex.
This fit (c.f.~figs.~2 and 3)
deviates appreciably from the other ones only at high energies
($\sqrt{s} >  $ 1.75~GeV).
The corresponding parameters are given in
the third row of table 1.

As a further test
of the model we have also calculated the cross section for
the process $\pi^- +p \rightarrow \eta+n$
\begin{equation}
\label{eq:i}
\sigma_{\pi\eta}
=\frac{4\pi}{q^2} |T _{\pi\eta}|^2.
\end{equation}
The $T$-matrix element $T _{\pi\eta}$ is approximated by its l=0 component
obtained in our model. In fig.~4 the resulting cross section
is compared with the experimental data
compiled in ref.~\cite{16}. We find good agreement in the energy range of the
S$_{11}$(1535) resonance. At higher energies contributions of higher lying
resonances become important (see below).

The S$_{11}$(1650) resonance, which was not included in previous models
\cite{15,2}, plays an important role both in elastic $\pi$N scattering and
$\eta$-production. The coupling strengths of one resonance are affected by
the presence of the other one, since the two resonances overlap due to their
relatively large widths.
At an isolated
resonance the real part generally goes through zero near the resonance energy.
The behaviour of the real part of the $T$-matrix in the S$_{11}${} channel
can only be described by
taking two overlapping resonances into account.
Furthermore, the upper resonance, although it couples only weakly to the
$\eta$-meson,
plays a crucial role in the determination of $g_{\eta\mbox{\scriptsize
N}\mbox{\scriptsize N}^*_1}$.
In order to demonstrate this, we have repeated the fit
in a reduced model, where the upper resonance is neglected. A reasonable
description of the data can be obtained for energies below $\sqrt{s}$ = 1.55
GeV, where one would expect only a small contribution from the
S$_{11}$(1650). However,
compared to the complete model we find a strong increase of both
coupling constants
($g_{\pi\mbox{\scriptsize N}\mbox{\scriptsize N}^*_1}$ = 0.80 and
$g_{\eta\mbox{\scriptsize N}\mbox{\scriptsize N}^*_1}$ = 2.38). Furthermore,
the $\eta$-production data cannot be described within the reduced model; at
its maximum the cross section is about 40 \% too large.

\section{Conclusions}
We have constructed a model for $\pi$N scattering in the elastic as well as
in the major inelastic S$_{11}$ channels. The parameters of the model are
determined by fitting only data for elastic scattering. Our best fit yields
$g_{\eta\mbox{\scriptsize N}\mbox{\scriptsize N}^*_1}$~=~2.10, which differs
considerably from the values obtained in most earlier calculations [5,10,21].
Vetter {\em et al.} \cite{6} find a value close to ours by fitting the
partial decay widths of the S$_{11}$ resonance.
However, we caution the reader that the coupling
constants discussed here, in particular $g_{\eta\mbox{\scriptsize
N}\mbox{\scriptsize N}^*_1}$, may be model dependent (see e.g. the discussion
at the end of section 4).

We find a total width of the S$_{11}$(1535) of 162~MeV
with branching ratios of 55$\,\%${} to $\eta$N, 41$\,\%${} to $\pi$N{}
and 4$\,\%${} to $\pi\pi$N, respectively. The S$_{11}$(1650) decays with
77$\,\%${}
to $\pi$N and with 23$\,\%${} to $\pi\pi$N{}, with a total width of 293~MeV.
The branching ratios of both resonances and the width of the S$_{11}$(1535)
are
consistent with the values given by the particle data group \cite{10}, while
the width of the S$_{11}$(1650) is larger. We note, however, that the
resonance
parameters, in particular the width, strongly depend upon the treatment
of the background.
For example, in the reduced model, where the upper resonance is
neglected, the total width of the S$_{11}$(1535) is 204 MeV.
The resonance parameters quoted in \cite{10} are
determined in a parametrization, which includes a phenomenological
background. In our model, this background is, at least in part, included in
the large width of the S$_{11}$(1650).

In order to explore how reliable our determination of
$g_{\eta\mbox{\scriptsize N}\mbox{\scriptsize N}^*_1}$ is, we have performed
constrained fits, where this coupling constant was fixed to the values 1.9
and 2.3. The quality of the fit to the elastic $T$-matrix is slightly worse
than in the optimal case. In the
figs.~2 and 3 these fits are almost indistinguishable from the original
fit. However, the cross section for $\eta$-production depends strongly on
the value of $g_{\eta\mbox{\scriptsize N}\mbox{\scriptsize N}^*_1}$. This is
illustrated by the band in fig.~4, where the upper edge corresponds to
$g_{\eta\mbox{\scriptsize N}\mbox{\scriptsize N}^*_1}$~=~2.3 and the lower
edge to $g_{\eta\mbox{\scriptsize N}\mbox{\scriptsize N}^*_1}$~=~1.9. Thus,
the optimal fit to the elastic $\pi$N-scattering $T$-matrix yields a good
description for the inelastic $\eta$-production channel {\it without any
further adjustment of parameters}.

At energies beyond the S$_{11}$(1535), the calculated
$\eta$-production cross section
is too low. The missing cross section is probably due to higher lying
resonances, not included in our model. The coupling to the P$_{11}$(1440)
and D$_{13}$(1520), which have been studied previously, seems to be rather
weak \cite{2}.
This is consistent with our model, where there is little room for additional
contributions at energies below 1.55~GeV. The next resonance which
couples
strongly to the $\eta$-meson \cite{10} is the P$_{11}$(1710). A rough estimate
of its contribution is obtained by using \cite{Jackson}
\begin{equation}
\label{uni}
\Delta \sigma_{\pi\eta}={4 \pi\over q^2}\left({2\over
3}\right){{1\over 4}\Gamma_{\pi{\mbox{\scriptsize
N}}} \Gamma_{\eta{\mbox{\scriptsize N}}}\over (\sqrt{s}-m_R)^2 +{1\over 4}
 (\Gamma_{tot})^2},
\end{equation}
where $m_R$ = 1.71~GeV is the resonance energy and $\Gamma_i$ and
$\Gamma_{tot}$ = 50-250~MeV are the partial and total widths of the
P$_{11}$(1710). We use $\Gamma_{tot}$ = 200 MeV, while for the branching ratios
we take the maximum values \cite{10} $\Gamma_\pi/\Gamma_{tot}$~=~0.2 and
$\Gamma_\eta/\Gamma_{tot}$~=~0.4. The energy dependence of the widths is
accounted for by multiplying $\Gamma_i$ and $\Gamma_{tot}$ by the
corresponding phase-space factors.
Adding this contribution to the cross section
yields the dash-dotted line in fig.~4.

In summary, we have performed a coupled-channel calculation of
$\pi$N scattering in the S$_{11}$ channel including Born diagrams and the two
resonances S$_{11}$(1535) and S$_{11}$(1650). The parameters of our model
were determined by fitting the $T$-matrix for elastic $\pi$N scattering to
empirical data. An excellent fit is obtained for all energies up to
$\sqrt{s}$ = 1.75 GeV. Furthermore, with the same parameters
the total cross section for the inelastic
process $\pi^-+p\rightarrow\eta+n${} is also well reproduced in
the region of the two resonances.

Our model can be used in calculations of other processes,
e.g. photoproduction of $\eta$-mesons or $\eta$-production
in hadronic and heavy-ion collisions. Work on the
photoproduction of $\eta$-mesons on protons is in progress. There the
present model is used to describe the strong-interaction part of the
$T$-matrix.

\section*{Acknowledgement}

We thank G.~H\"ohler, M.~Soyeur and W.~Weise for valuable discussions.
\vspace{3cm}
\section*{Figures}
\begin{center}
\parbox{13cm} {
Fig.~1: $K$-matrix elements for the three open channels,
with $i= \eta,\zeta$.}

\vspace{1cm}

\parbox{13cm}{
Fig.~2: Imaginary part of the $T$-matrix amplitude
for $\pi$N scattering in the elastic S$_{11}$ channel as a
function of the c.m.~energy.
The data are taken from the Karlsruhe-Helsinki analysis [11],
the solid line is our best fit and the dash-dotted line
shows our fit without a form factor at the $\pi$NN vertex.}

\vspace{1cm}

\parbox{13cm}{
Fig.~3: Same as fig.~2 for the real part of the $T$-matrix.}

\vspace{1cm}

\parbox{13cm}{
Fig.~4: Total cross section for the process $\pi^- + p \rightarrow
\eta + n${} as a function of the laboratory pion momentum.
The data are from ref.~20. The heavy solid line is computed with
the parameters of our best fit
to the elastic $\pi$N-scattering data. The boundaries of the
shaded area correspond to $\pm 10\%${} deviations of
$g_{\eta \mbox{\scriptsize N}\mbox{\scriptsize N}^*_1}$ from its
optimal value (for details see text). The dash-dotted line includes the
contribution from the P$_{11}$(1710) resonance. The arrows indicate
the location of the S$_{11}$(1535) and P$_{11}$(1710) resonances. }
\end{center}


\newpage
\begin{thebibliography}{11a}
\bibitem{1}{B. Krusche, preprint, to be published in {\sl Proc. II TAPS
Workshop}, Alicante, 1993}
\bibitem{1a}{B. Schoch, Acta Phys. Polonica {\bf24} (1993) 1765}
\bibitem{1b}{E. Chiavassa {\sl et al.}, preprint 1994}
\bibitem{1c}{F.-D. Berg {\sl et al.} Z. Phys. {\bf A340} (1991) 297; Phys.
Rev. Lett. {\bf 72} (1994) 977}
\bibitem{2}{C. Bennhold, H. Tanabe, Nucl. Phys. {\bf A530} (1991)
625}
\bibitem{3}{M. Benmerrouche and N. C. Mukhopadhyay, Phys. Rev. Lett.
{\bf 67} (1991) 1070}
\bibitem{4}{J. F. Germond and C. Wilkin, Nucl. Phys. {\bf A518}
(1990) 308}
\bibitem{5}{J. M. Laget, F. Wellers, J. F. Lecolley, Phys. Lett.
{\bf B257} (1991) 254}
\bibitem{6}{T. Vetter, A. Engel, T. Biro and U.Mosel, Phys.
Lett. {\bf B263} (1991) 153}
\bibitem{7}{H. C. Chiang, E. Oset, L.C. Liu, Phys. Rev.
{\bf C44} (1991) 738}
\bibitem{8}{G. H\"{o}hler, {\sl Pion-Nucleon Scattering},
Landolt-Bornstein Vol.I/9b2, (Springer, Berlin, 1983) and private
communication}
\bibitem{9}{R. E. Cutkosky, Phys. Rev. {\bf D20} (1979) 2804 and 2839}
\bibitem{10}{Particle Data Group, {\sl Review of Particle Properties},
Phys. Rev. {\bf D45} (1992)}
\bibitem{12}{B. Lee, {\sl Chiral Dynamics}, (Gordon and Breach, New York,
1972)}
\bibitem{13}{D. Campbell, Proceedings Les Houches 1977,
{\sl Nuclear Physics with heavy ions and mesons} (North-Holland, Amsterdam,
1978)}
\bibitem{11}{W. Grein and P. Kroll, Nucl. Phys. {\bf A338} (1980) 332}
\bibitem{11a}{C. J. Joachain, {\sl Quantum Collision Theory},
(North-Holland, Amsterdam, 1975)}
\bibitem{4a}{G. E. Brown and W. Weise, Phys. Rep. {\bf 22} (1975) 279 \\
E. Oset, H. Toki and W. Weise, Phys. Rep. {\bf 83} (1982) 282 \\
T. Ericson and W. Weise, {\sl Pions and Nuclei},
(Clarendon Press, Oxford, 1988)}
\bibitem{14}{B. C. Pearce and B. K. Jennings, Nucl. Phys.
{\bf A528} (1991) 655}
\bibitem{16}{M. Clajus, B. M. K.
Nefkens in $\pi$N-Newsletter No.7, ed. R. E. Cutkosky,
G. H\"ohler, W. Kluge, B. M. K. Nefkens, IEKP University of Karlsruhe and
Department of Physics UCLA, Dec. 1992}
\bibitem{15}{R. S. Bhalerao und L. C. Liu, Phys. Rev. Lett. {\bf 54}
(1985) 865 }
\bibitem{Jackson} {J.D. Jackson, Nuovo Cimento {\bf 34} (1964) 1645}
\end{thebibliography}
\end{document}